\documentstyle[prb,aps,epsf]{revtex}
\begin{document}

\draft

\twocolumn[\hsize\textwidth\columnwidth\hsize\csname@twocolumnfalse%
\endcsname

\title{PHASE DIAGRAM OF ADSORBATE-INDUCED ROW-TYPE-ALIGNMENTS}
\author{M.~KANG$^*$, K.~YASUTANI$^{\dagger}$, and 
M.~KABURAGI$^{\dagger,\ddagger}$}
\address{$^*$ {Kansai University of International Studies, }\\
{Shijimi, Miki 673-0521, Japan}\\
{$^{\dagger}$ { Graduate School of Science and Technology, Kobe University}}\\
{Rokkodai, ~Nada, ~Kobe ~657-8501, Japan}\\
{$^{\ddagger}$ { Faculty of Cross-Cultural-Studies, Kobe University}}\\
{ Tsurukabuto, ~Nada, ~Kobe ~657-8501, Japan}\\
}

\maketitle

\begin{abstract}
The phase diagram of adsorbate-induced row-type-alignments, such as 
missing-row reconstructions induced by adsorbate-atoms on the FCC(110) 
surface, is calculated by the Blume-Emmery-Griffiths (BEG) model. In 
the model, we introduce adatom-adatom and dipole-dipole interactions 
between nearest-neighbor (NN) and next-nearest-neighbor (NNN) rows. 
The calculation of the temperature versus adatom chemical potential 
phase diagram is performed using mean-field approximation.
It is indicated that when  NN and  NNN interactions are competitive, 
there appear either dipole or coverage modulated  (incommensurate)  
phases at high temperatures for wide regime of the interactions.
\end{abstract}
\pacs{~}
]
\section{Introduction}
It is well known that row-type-alignments are induced in many cases 
on transition metal surfaces by adsorbates.  A typical example is the 
(1 $\times$ $n$) missing-row reconstruction (MRR) of FCC(110) metal surface 
covered by adatoms, such as H/Cu(110) {\cite{H/Cu}}, 
O/Rh(110) {\cite{O/Rh1,O/Rh2}}, and
O/Pd(110) {\cite{O/Pd1,O/Pd2}}. 
In these systems, the rows of the the metal atoms in [1$\bar{1}$0] directions 
are accompanied by zigzag chains of adatoms sitting in the threefold 
FCC hollow sites on the [111] microfacets of the surface and each $n$th
row is missing. It should be noted  the arrangement of the missing 
row at low coverage $c$ of adatoms suggests that the interactions 
between the rows of the metal atoms are of repulsive.

On the theoretical side,  P.~J.~Kundrotas {\it et al\/} {\cite{KLR}} 
recently showed that the 2 dimensional Blume-Emmery-Griffiths(BEG)-like 
model {\cite{BEG}} is a useful model to analyze these ordering 
phenomena of row-type-alignments including the adatom-metal dipoles. 
Their excellent calculation of the finite temperature phase diagram for 
O/Rh(110) is, however, restricted to the case where the 
nearest-neighbor (NN) interactions along [001] direction is much 
stronger than the next-nearest-neighbor (NNN) interactions. 
 
The purpose of this paper is to extend the analysis, including the ground
state analysis, to the case of the relatively strong NNN interactions.
In the case of axial-next-nearest-neighbor-Ising (ANNNI) 
model{\cite{ANNNI}}, when  NN and  NNN interactions are competitive, 
an introduction of the relatively strong NNN interactions usually 
give rise to change of the ground state structures and also induces 
entropic modulated phases at high temperatures. 
From this fact, we may expect that the BEG model will also exhibit 
a rich variety of phases due to the frustration of the interactions. 

In this paper, after establishing  our model, we determine the 
ground state phase diagram by energy comparison method.  This ground 
state analysis shows the new structure corresponding to the 
$\langle uudd \rangle$ structure of the ANNNI model appears under 
the certain conditions for  
NNN interactions. We then derive the mean-field equations to investigate 
qualitative feature of the finite temperature phase diagrams. 
Solving the linearized mean-field equations analytically,  we obtain 
the condition for the interactions to realize the stable 
modulated phases. Numerical analysis of the mean-field equations 
is performed to calculate the finite temperature phase diagrams for 
several interaction regimes, including the region satisfying 
the modulated phase conditions. These analyses show that 
there appear either dipole or coverage modulated    
phases at high temperatures for certain regime of the interactions.

\newcounter{tmpsec}
\addtocounter{tmpsec}{1}
Organization of this paper is as follows:  
In the next section (section 
\addtocounter{tmpsec}{1}
\Roman{tmpsec}
), 
after setting up the model for the MRR ordering, we construct 
the ground state phase diagram. 
Section 
\addtocounter{tmpsec}{1}
\Roman{tmpsec} 
deals with derivation of the mean-field equations and 
solution corresponding to the modulated phase. Section 
\addtocounter{tmpsec}{1}
\Roman{tmpsec} 
is devoted to the numerical analysis. 
Finally, summary and concluding remarks are given in 
section 
\addtocounter{tmpsec}{1}
\Roman{tmpsec} 
.

\section{Model and Ground State Analysis}
It is plausible to expect that the clusterlike bonding between 
metal atom and adatom produces a dipole moment by transfer of electron.
In case of zigzag adsorption, there exists the asymmetric bonding 
between metal atom and adatom.
The asymmetry can be theoretically treated by the dipole
moments with opposite directions, say "up" or "down" dipoles.
Therefore, a site of the surface where metal atom sits in is in 
one of three states, missing of metal atom (vacant), occupied 
by metal atom  with "up" or "down" dipoles.

In order to analyze the ordering phenomena of row-type-alignments 
including the adatom-metal dipoles,  we introduce pseudo "spin" variable 
$\sigma_{i}$ at ${i}$ site to represent the three states of the site ${i}$,  
vacant ($\sigma_{i}\,=\,0$), occupied by metal atom  with "up" or "down" 
dipoles ($\sigma_{i}\,=\,\pm 1$). 
Since the occupation number of metal atom (accompanied by adatom) 
at $i$ site is given by $\sigma_{i}^2$, 
the Hamiltonian of the system is represented by the BEG model as
%
%
\begin{eqnarray}
{\cal H}  &=  \displaystyle{\sum_{ij}}J_{ij}\sigma_{i}\sigma_{j} +
     \displaystyle{\sum_{ij}}K_{ij}(\sigma_{i}^2 - 2/3)(\sigma_{j}^2 - 2/3) \nonumber\\
 &~+ \mu \displaystyle{\sum_{i}}(\sigma_{i}^2 - 2/3)
\label{eq:Hami1}
\end{eqnarray}
where $J_{ij}$ ($K_{ij}$) is interaction between $ij$ pair of the dipoles 
(metal atoms), $\mu$ is the chemical potential. 
$\sigma_i^2 = 1(0)$ means a metal atom occupies the surface site $i$ (or not).
Here we use 
$(\sigma_{i}^2 - 2/3)$ instead of $\sigma_{i}^2$ for convenience 
of the later calculation. 
Following  P.~J.~Kundrotas {\it et al\/} {\cite{KLR}}, we assume that 
\begin{equation}
 (J_{ij}, K_{ij}) = \left\{
                      \begin{array}{@{\,}ll}
                      {(J_0, K_0)} & \mbox{${ij}=$ {NN pair in [1$\bar{1}$0] 
direction} }\\
                      {(J_1, K_1)} & \mbox{${ij}=$  {NN pair in [001] 
direction}  }\\
                      {(J_2, K_2)} & \mbox{${ij}=$  {NNN pair in [001] 
direction}  }\\
                      {(0, 0)} & \mbox{otherwise
  }
                      \end{array}
                      \right. 
\end{equation}
and farther neighbor interactions are negligible. Then the Hamiltonian is rewritten 
as
\begin{eqnarray}
{\cal H}  &=  \displaystyle{\sum_{\ell,m}} \Bigl\{
     J_{0}\sigma_{\ell\,m}\sigma_{\ell\,m+1} +
     K_{0}(\sigma_{\ell\,m}^2 - 2/3)(\sigma_{\ell\,m+1}^2 - 2/3) \nonumber \\ 
&~+ ~J_{1}\sigma_{\ell\,m} \sigma_{\ell+1\,m} + 
     K_{1}(\sigma_{\ell\,m}^2 - 2/3)(\sigma_{\ell+1\,m}^2 - 2/3) \nonumber \\
&~+ ~J_{2}\sigma_{\ell\,m} \sigma_{\ell+2\,m} + 
     K_{2}(\sigma_{\ell\,m}^2 - 2/3)(\sigma_{\ell+2\,m}^2 - 2/3) \nonumber 
\end{eqnarray}
\vspace{-1.0cm}
\begin{equation}
~+ ~ \mu (\sigma_{\ell\,m}^2 - 2/3)\Bigr\}~, 
\label{eq:Hami2}
\end{equation}
where $\ell$ ($m$) refers site-number along [001] ( [1$\bar{1}$0]), 
namely $\ell$- ($m$-), direction.

In the MRR-type-alignments, metal atoms form a chain-like structure and
the adatoms are ordered in zigzag arrangement which corresponds to 
"antiferromagnetic" ordering of the $\sigma_{\ell\,m}$ along $m$-direction.
From these experimental facts, the interactions in Eq.(\ref{eq:Hami2}) 
should satisfy the
\begin{equation}
   J_0 > 0~, ~~K_0 < 0~, ~~K_1 > 0~, ~~K_2 > 0~.
\label{eq:J0K0}
\end{equation}
It should be noted that the Hamiltonian (\ref{eq:Hami2}) can be 
transformed into "ferromagnetic" ($ J_0 < 0$) one by  the transformation
$   \sigma_{\ell\,m} \rightarrow (-1)^{m}\sigma_{\ell\,m}$ .

Now we determine the ground state phase diagram by comparing the 
energies of several  "spin" arrangements. From the condition
$J_0 > 0$ and $K_0 < 0~$, the "spin" arrangement in the ground state
is expected to be represented by $m$-independent "spin" 
${\hat{\sigma}}_{\ell}$ as
\begin{equation}
\sigma_{\ell\,m} = (-1)^{m}{\hat{\sigma}}_{\ell}.
\label{eq:mtrans2}
\end{equation}
To represent the "spin" arrangement, we introduce the notation 
which indicates the repeating unit of ${\hat{\sigma}}_{\ell}$ as
$\langle {\hat{\sigma}}_{1} {\hat{\sigma}}_{2} \cdot\cdot\cdot 
{\hat{\sigma}}_{n}\rangle$; for example $\langle  ud0 \rangle$ 
represents the 
"spin" 

\begin{figure}[t]
\begin{center}
\noindent
\epsfxsize=0.44\textwidth
\epsfbox{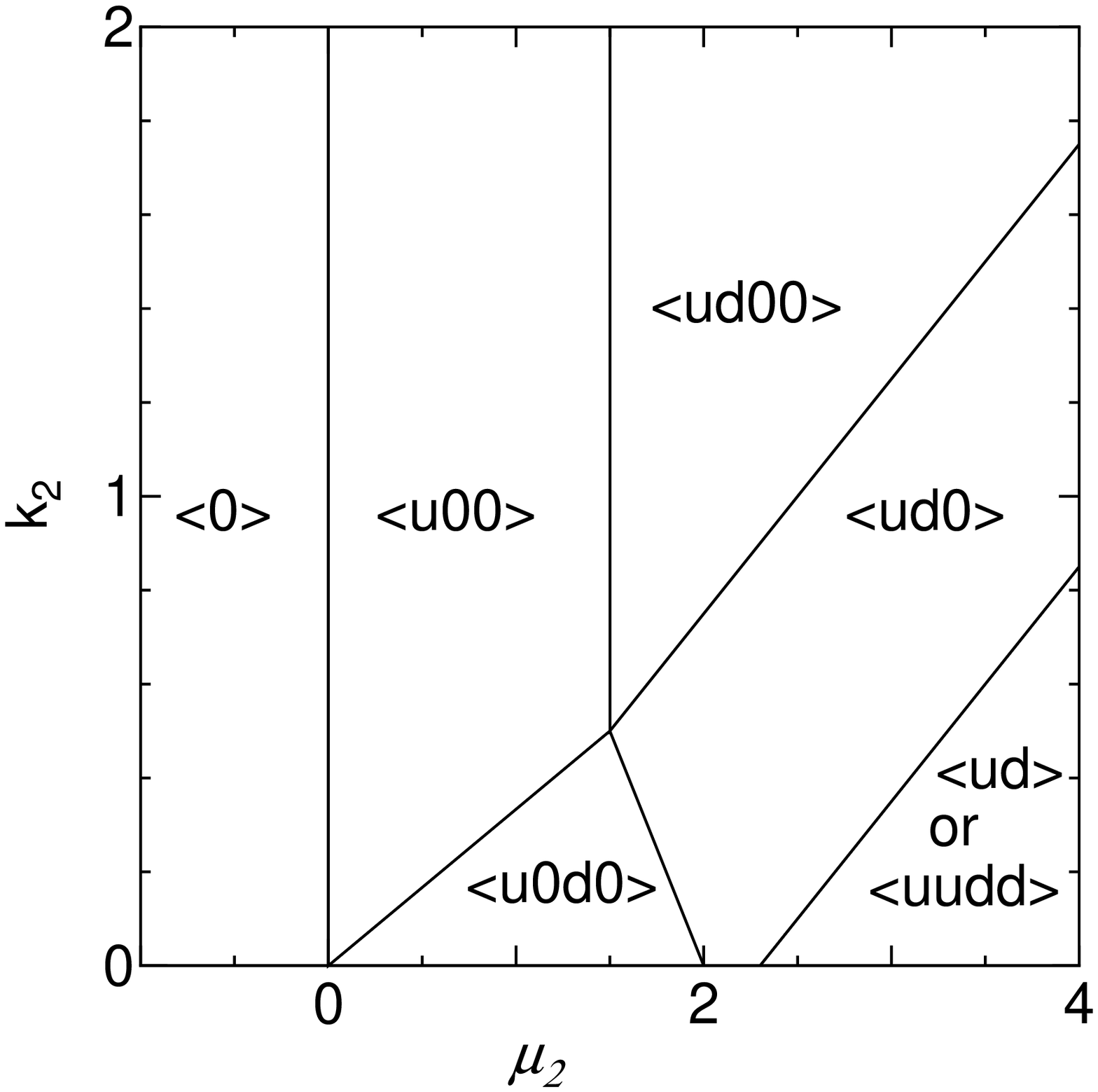}
\end{center}
\vspace{-0.5cm}
\begin{center}
\noindent
\epsfxsize=0.44\textwidth
\epsfbox{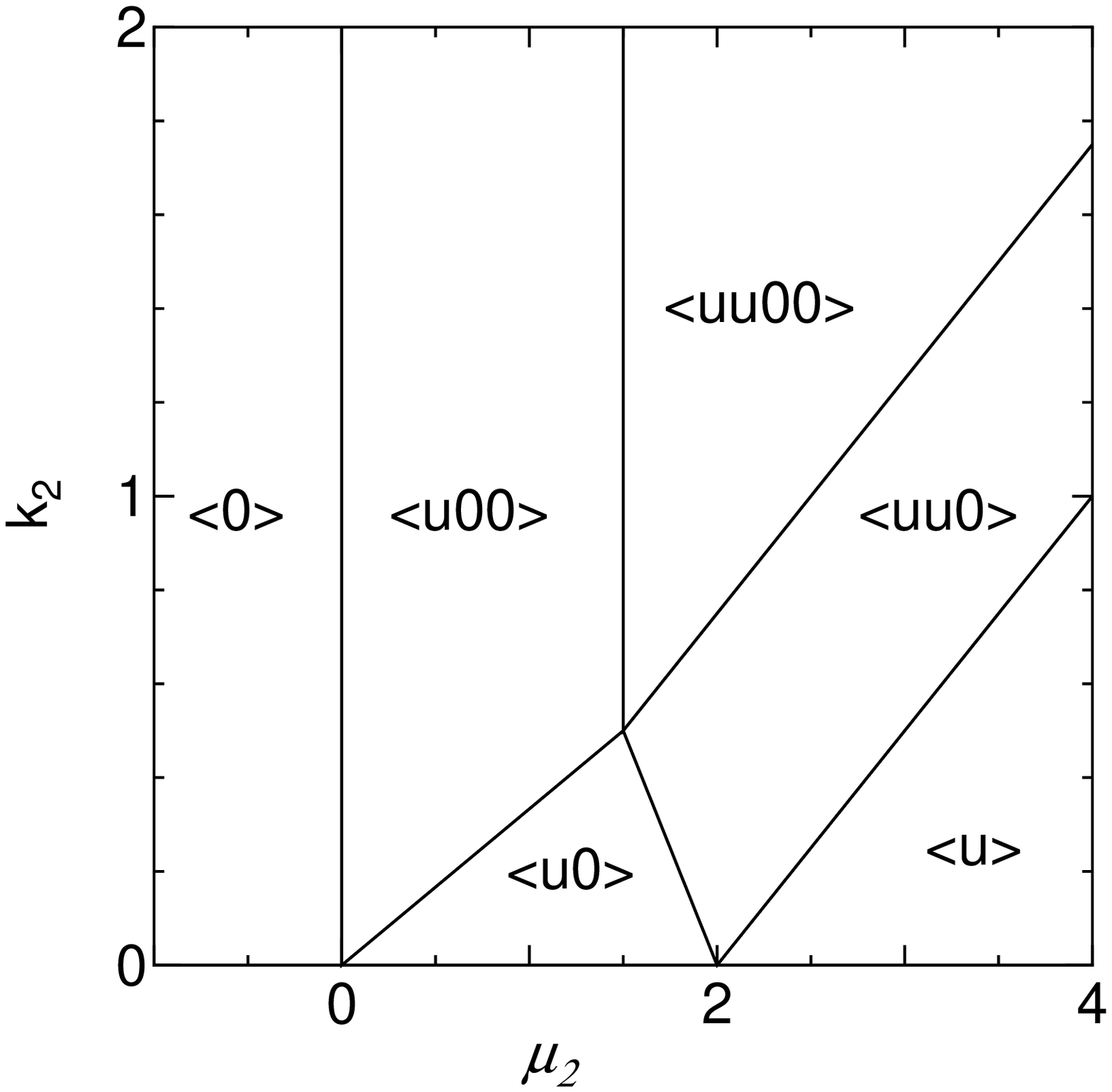}
\end{center}
\caption{Ground state phase diagrams for $K_0 < 0$, $J_0 >0$ 
and $K_i > |J_i| (i=1, 2) $
in $k_2$-$\mu_2$ plane, where
$k_2$=$(K_2-|J_2|)/(K_1-|J_1|)$ 
and 
$\mu_2$=$\bigl(-\mu +{\frac{4}{3}}(K_0+K_1+K_2)$ 
$+|J_0|-K_0\bigr)/(K_1-|J_1|)$: 
(a:upper) for $J_1 >0$, $J_2 >0$; (b:lower) for $J_1 <0$, $J_2 <0$.
Boundary lines in the diagrams are described as:
$\mu_2=0$ for $\langle 0\rangle$ and $\langle u00\rangle$; 
$\mu_2=3k_2$ for
$\langle u00\rangle$ and $\langle u0d0\rangle$ in (a), 
for $\langle u00\rangle$ and $\langle u0\rangle$ in (b);
$\mu_2=3/2$ for
$\langle u00\rangle$ and $\langle ud00\rangle$ in (a), 
for $\langle u00\rangle$ and $\langle uu0\rangle$ in (b);
$\mu_2=2-k_2$ for
$\langle u0d0\rangle$ and $\langle ud0\rangle$ in (a), 
for $\langle u0\rangle$ and $\langle uu0\rangle$ in (b);
$k_2=\mu_2/2-1/4$ for
$\langle ud00\rangle$ and $\langle ud0\rangle$ in (a), 
for $\langle uu00\rangle$ and $\langle uu0\rangle$ in (b);
$k_2=(\mu_2-\Delta_j)/2$ for
$\langle ud0\rangle$ and $\langle ud\rangle$ or 
$\langle uudd\rangle$ in (a) where $\Delta_j= 
Min\Bigl(6J_2/(K_1-|J_1|), 3J_1/(K_1-|J_1|)\Bigr)$; $k_2=(\mu_2-2)/2$
 for $\langle uu0\rangle$ and $\langle u\rangle$ in (b).
}
\end{figure}

\noindent
arrangement with repeating unit 
$({\hat{\sigma}}_{1}{\hat{\sigma}}_{2}{\hat{\sigma}}_{3})\,=\,(+1~-1~0)$.
The target "spin" arrangements of the energy comparison are

\noindent
Case A: $ J_1 > 0~, ~~J_2 > 0~$

$\langle 0 \rangle ~, ~~\langle u00 \rangle ~, ~~\langle u0d0 \rangle ~,
~~\langle ud00 \rangle ~, ~~\langle ud0 \rangle ~, 
~~\langle ud \rangle ~, ~~\langle uudd \rangle 
$

\noindent
Case B: $ J_1 < 0~, ~~J_2 < 0~$

$\langle 0 \rangle ~, ~~\langle u00 \rangle ~, ~~\langle u0 \rangle ~,
~~\langle uu00 \rangle ~, ~~\langle uu0 \rangle ~, ~~\langle u \rangle ~$

\noindent
It is noticed that the results for the case C : $ J_1 < 0~, ~~J_2 > 0~$
(case D :$ J_1 > 0~, ~~J_2 < 0~$ ) can be easily obtained from the
results for case A (case B) by the transformation 
$ \hat{\sigma}_{\ell} \rightarrow (-1)^{\ell}{\hat{\sigma}}_{\ell}$.

Figure~1 shows the ground state phase diagram in  $\mu_2$-$k_2$ plane 
obtained under the condition 
$   K_i > |J_i|\, (i=1, 2) $, where  $\mu_2$ and $k_2$  are given as  
\begin{eqnarray}
k_2 =&(K_2-|J_2|)/(K_1-|J_1|)\, ,\\
\mu_2=&\bigl(-\mu +{\frac{4}{3}}(K_0+K_1+K_2)+ \nonumber \\
&~|J_0|-K_0\bigr)/(K_1-|J_1|)\,.
\end{eqnarray}
For large $\mu_2$ regime in Fig.1(a), 
either $\langle ud \rangle$ or $\langle uudd \rangle$ is realized 
depending upon $2J_2 <J_1$ or $2J_2 >J_1$, respectively. 

For large $K_2$, the energetically preferable structure may
be the one containing no $K_2$ bonding, such as $\langle uu00 \rangle$ or 
$\langle ud00 \rangle$. Actually, as is shown in Fig.~1, 
there appears either  $\langle uu00 \rangle$ or 
$\langle ud00 \rangle$ depending sign of $J_1$ for the relatively large 
$K_2$. Therefore we might observe the different sequences of the MRR 
with increasing the chemical potential if 
we prepare the surface with relatively large $K_2$.
It is remarked here that there exist many structures degenerate in energy 
in the region of the phase containing two adjacent "0". For example, 
structures consisting of any combination of 
$\langle u00 \rangle $ ($\langle ud00 \rangle$) and 
$\langle d00 \rangle $ ($\langle du00 \rangle$) are degenerate in 
energy with the structure
$\langle u00 \rangle $ ($\langle ud00 \rangle$).

\section{Mean-Field Analysis and Modulated Phases}
In this section we derive the mean-field equations to construct 
phase diagrams at finite temperatures. 
The mean-field energy $\epsilon$ per $m$-direction line is
obtained as
\begin{eqnarray}
{\epsilon}  =&  \displaystyle{\sum_{\ell}} \Bigl\{
     -|J_{0}|x_{\ell}^2 + \frac{K_{0}}{9}(y_{\ell}^2+2) \nonumber \\
&~+ \displaystyle{ J_{1}x_{\ell} x_{\ell+1} 
+ \frac{K_{1}}{9}(y_{\ell}y_{\ell+1}+2)
} \nonumber \\
&~+ \displaystyle{ J_{2}x_{\ell} x_{\ell+2} + 
                    \frac{K_{2}}{9}(y_{\ell}y_{\ell+2}+2) }
-   \frac{\mu}{3}y_{\ell}
\label{eq:Eng2}
\end{eqnarray}
where $x_{\ell}$ and $y_{\ell}$ are $m$ independent thermal average of  
$(-1)^m \sigma_{\ell\,m}$ and $(2 - 3\sigma_{\ell\,m}^2)$, respectively.
Since $\sigma_{\ell\,m}^2$ is the occupation 
number of metal atom (accompanied by adatom) at $\ell\,m$ site, 
as mentioned in previous section,  $c_{\ell}=(2 -y_{\ell})/3$ 
gives the $m$ independent thermal average of 
atomic concentration of metal or adatom (coverage) at  $\ell\,m$ site.

The mean-field entropy $s$ per  $m$-direction line can be expressed
in a unit of $k_B=1$ by $x_{\ell}$ and $y_{\ell}$ as 
\begin{equation}
 s = -\displaystyle{\sum_{\ell} \sum_{\sigma}} R_{\ell}(\sigma)
\bigl\{\log R_{\ell}(\sigma) -1 \bigr\}~.
\label{eq:Ent3}
\end{equation}
where  $R_{\ell}(\sigma)$ is the $m$ independent probability of 
finding the $\ell\,m$ site at state $\sigma$ given as 
\begin{equation}
 R_{\ell}(\sigma) = {1 \over 3}\bigl(1+3\sigma x_{\ell}/2 +
(2-3\sigma^2)y_{\ell}/2\bigr).
\label{eq:Ent4}
\end{equation}

Optimization of the free energy $f\,=\,\epsilon-Ts$ per line
with respect to  $x_{\ell}$ and $y_{\ell}$ leads  the
so called mean-field equations as
\begin{eqnarray}
&~-2|J_{0}|x_{\ell}+J_{1}(x_{\ell-1}+ x_{\ell+1})  
   + ~J_{2}(x_{\ell-2}+ x_{\ell+2}) \nonumber \\
&~+ T\displaystyle{\sum_{\sigma}}{\sigma \over 2}\log R_{\ell}(\sigma)=0~,
\label{eq:MFEx}\\
&\displaystyle{
\frac{2K_{0}}{9} y_{\ell}~+ \frac{K_{1}}{9}(y_{\ell-1}+y_{\ell+1})
+   \frac{K_{2}}{ 9}(y_{\ell-2}+y_{\ell+2}+2)
} \nonumber\\
&\displaystyle{~-  ~ \frac{\mu}{3}
~+~T\sum_{\sigma}\frac{(2-3\sigma^2)}{6} \log R_{\ell}(\sigma)=0
}~.
\label{eq:MFEy}
\end{eqnarray}
where $T$ is the temperature.

When $\mu\,=\,0$, we easily obtain a stable "paramagnetic" solution 
$x_{\ell}\,=\,0$ and $y_{\ell}\,=\,0$ with coverage 
$c\,=\,2/3$ at high temperatures. To determine the transition temperature 
from the "paramagnetic" phase and the phase just below the transition 
temperature,  we linearize  the mean-field equations. Substitution 
of $x_{\ell}\,=\,X_q \exp (iq\ell)$ and $y_{\ell}\,=\,Y_q \exp (iq\ell)$ 
in the linearize equations leads us to the transition temperature at 
$\mu\,=\,0$ as maximum of 
\begin{eqnarray}
   T_{x}(q)= \displaystyle{{4 \over 3}}(|J_{0}|-J_{1}\cos(q) - ~J_{2}\cos(2q))~,
\label{eq:Tx}\\
   T_{y}(q)= \displaystyle{{4 \over 9}}(-K_{0} -K_{1}\cos(q) - ~K_{2}\cos(2q))~.
\label{eq:Ty}
\end{eqnarray}

When NNN interactions $|J_{2}|$~, $|K_{2}|$ are very small, either $q\,=\,0$ or
$q\,=\,\pi$ gives the maximum and there appears no modulated phases.
On the other hand, if NNN interactions satisfy the condition 
\begin{equation}
   |J_{2}/J_{1}| > 1/4 (J_{2} > 0) ~~~or~~~ |K_{2}/K_{1}| > 1/4 (K_{2} > 0)
\label{eq:Qcond}
\end{equation}
the system just below the transition temperature exhibits the modulated phase 
characterize by  wave number $\cos(Q)=-J_{1}/4J_{2}$ (dipole modulation)
or $\cos(Q)=-K_{1}/4K_{2}$ (coverage modulation).

\section{Numerical Results}
Since it is difficult to solve the mean-field equations analytically, 
we have to solve the equations numerically to obtain the phase diagrams.
In the present paper, we confine ourselves to the case of "ferromagnetic" 
interactions between rows, {\it  i.e } $J_1 <0$ and $J_2 <0$.
To demonstrate the change of ground state and appearance of the modulated phase, 
we deal with the three cases of interaction regime, 

\noindent
a) $J_0\,=\,1$, $K_0\,=\,-3$, $J_1\,=\,-1$, $K_1\,=\,3$, $J_2\,=\,-0.05$, 
and $K_2\,=\,0.5$~,

\begin{figure}[ht]
\begin{center}
\noindent
\epsfxsize=0.45\textwidth
\epsfbox{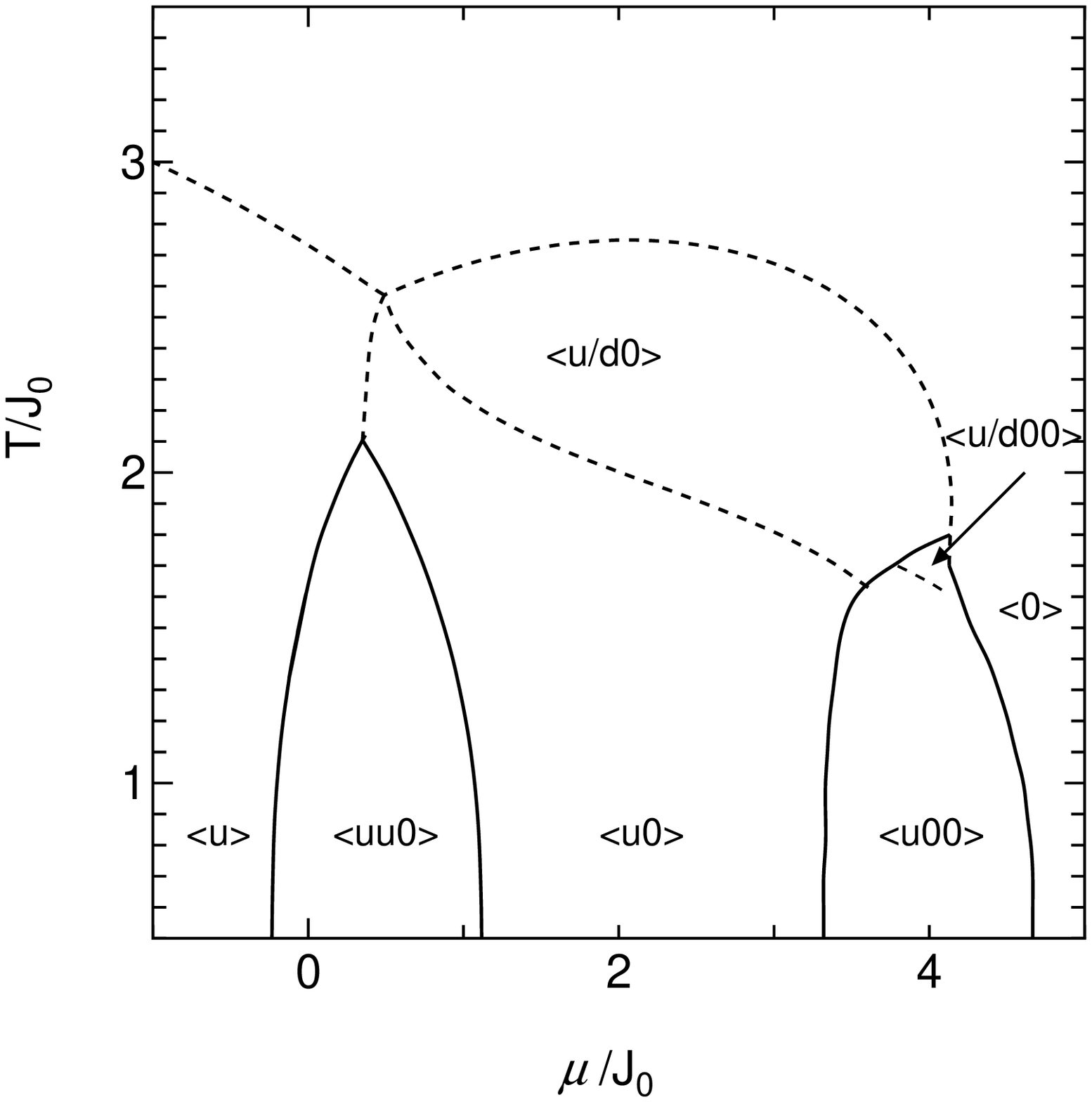}
\end{center}
\end{figure}

\vspace{-1.0cm}
\begin{figure}
\begin{center}
\noindent
\epsfxsize=0.45\textwidth
\epsfbox{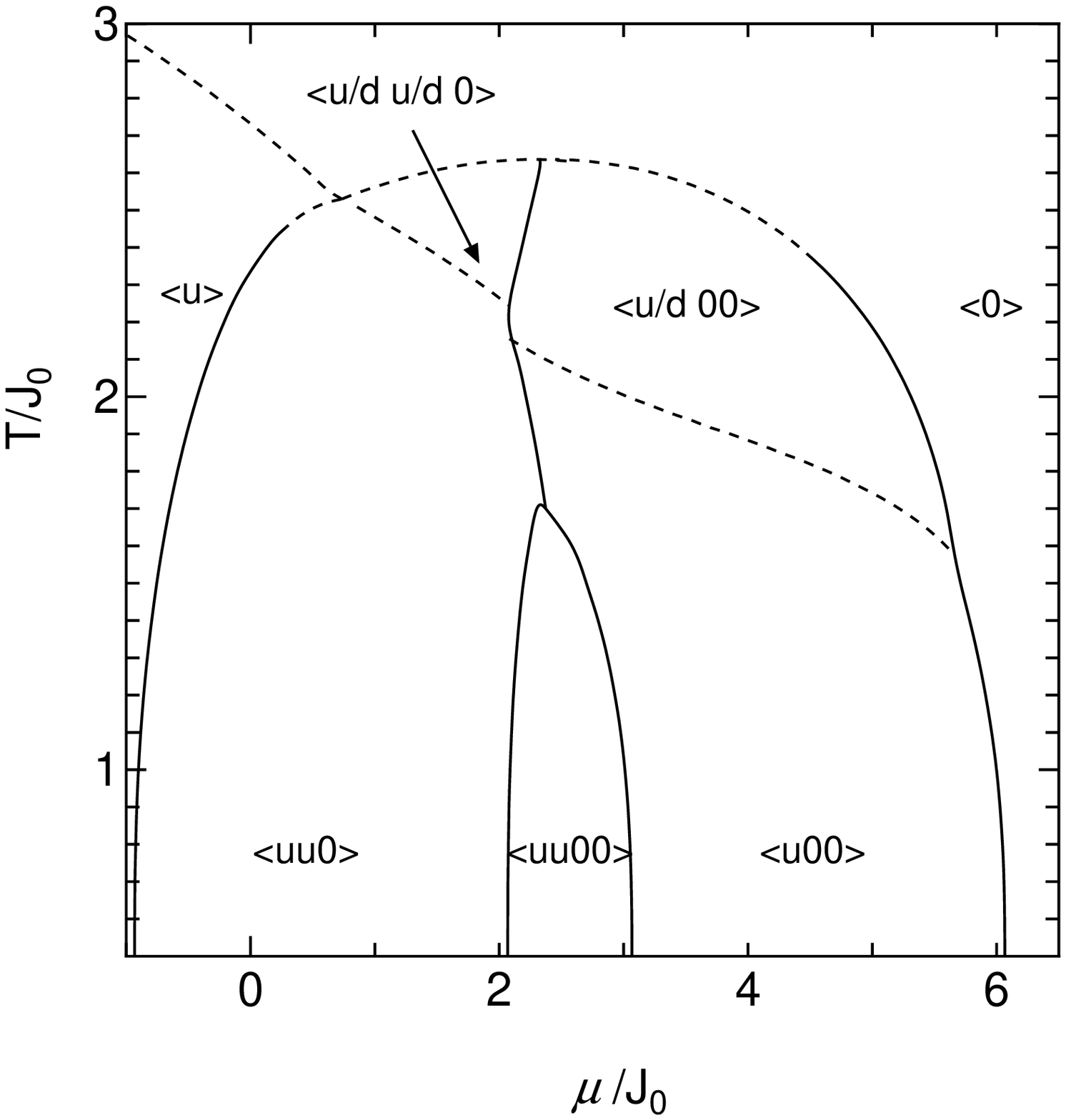}
\end{center}
\end{figure}

\noindent
b) $J_0\,=\,1$, $K_0\,=\,-3$, $J_1\,=\,-1$, $K_1\,=\,3$, $J_2\,=\,-0.05$, 
and $K_2\,=\,1.55$~,

\noindent
c) $J_0\,=\,1$, $K_0\,=\,-3$, $J_1\,=\,-.5$, $K_1\,=\,3$, $J_2\,=\,-0.03$, 
and $K_2\,=\,0.88$~.

\noindent
The interaction constants in the case a) are the same as Ref.{\cite{KLR}};
those in the case b) correspond to the regime $k_2\,=\,0.75 (>.5)$ where the 
$\langle uu00 \rangle$ appears as ground state (see Fig.1(b)); 
those in the case c) satisfy the condition $ K_{2}/K_{1}(\,=\,0.29333) > 
1/4 (K_{2} > 0)$, the modulated phase is expected to exist.

To construct the phase diagrams for the three cases of interaction 
regime at finite temperatures, 
we solve the mean-field equations within repeating unit 12, 
{\it  i.e } $(x_{12+\ell},y_{12+\ell})=(x_{\ell},y_{\ell})$ 
(${\ell}=1,2,\cdots,12$), for various

\begin{figure}
\begin{center}
\noindent
\epsfxsize=0.43\textwidth
\epsfbox{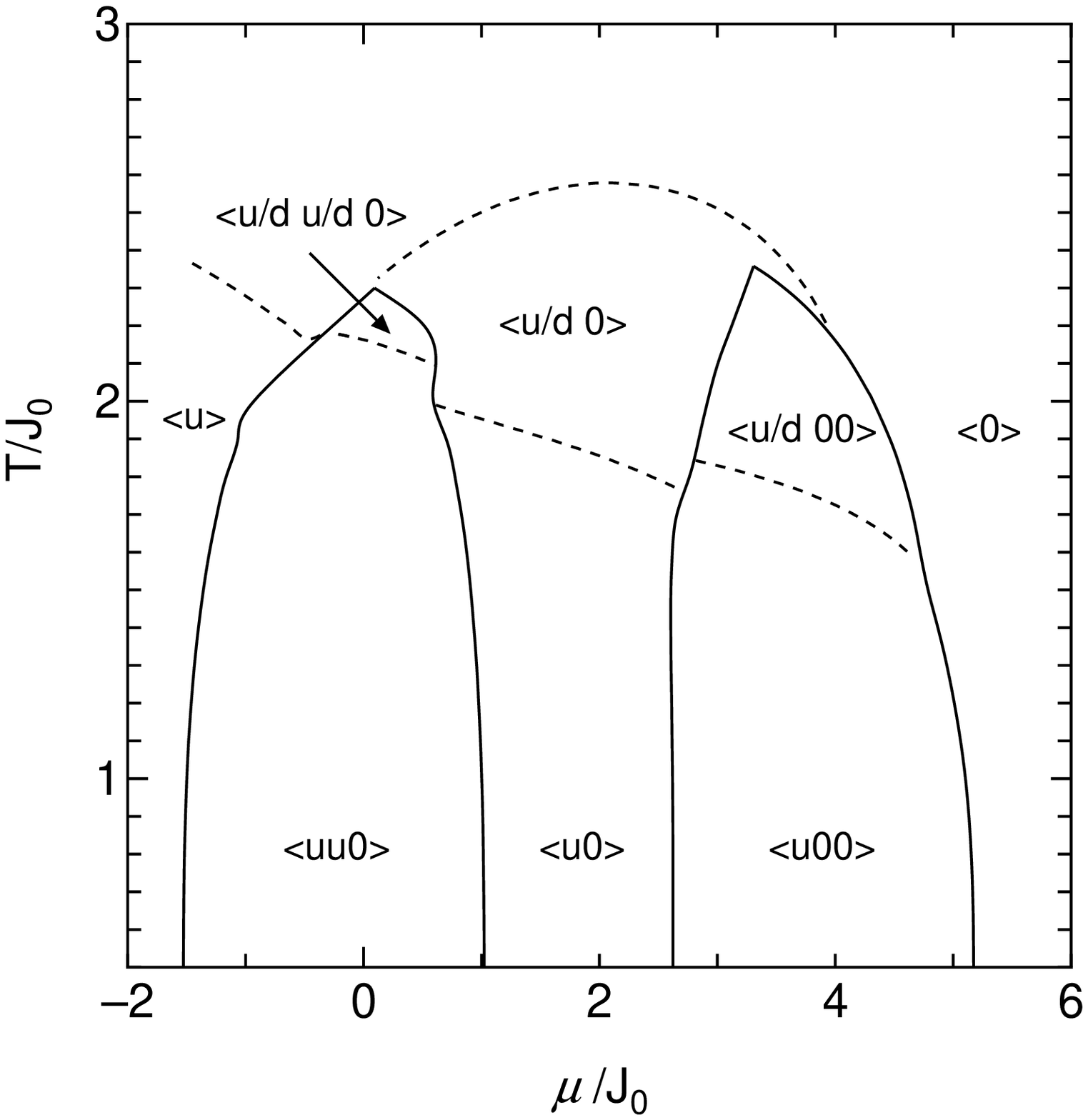}
\end{center}
\caption{
 $T$ versus $\mu$ phase diagrams:
(a:left upper) for $J_0=1$, $K_0=-3$, $J_1=-1$, $K_1=3$, 
$J_2=-0.05$, 
and $K_2=0.5$~;
(b:left lower) for $J_0=1$, $K_0=-3$, $J_1=-1$, 
$K_1=3$, $J_2=-0.05$,
and $K_2=1.55$~;
(c:right) for $J_0=1$, $K_0=-3$, $J_1=-0.5$, 
$K_1=3$, $J_2=-0.03$, 
and $K_2=0.88$~. The dotted(full) lines represent the 2nd(1st)
order phase transition lines. 
}
\end{figure}

\noindent
values of temperatures 
and chemical 
potential $\mu_2$ in Fig.1(b) along the line $k_2=0.225$ ( case a)), 
the line $k_2=0.75$ (case b)) and the line $k_2=0.34$ (case c)).

The calculated $T$ versus $\mu$ phase diagrams for the cases a), b), and 
c) are shown in Fig. 2(a), (b), and (c), respectively.  In the figures, 
$u/d$ denotes the "paramagnetic" row with $x\,=\,0$ and $y\,\neq\,0$.

It is natural to expect that at low temperatures, 
there appear successively the phases corresponding to
the ground state structures with increasing chemical potential; 
at high temperatures the rows prefer the "paramagnetic" state to 
"magnetic" state to gain free energy by entropy term.
As seen in Figs.2(a) to (c), all of our phase diagrams show this tendency.

It is clear that the phase diagram in Fig.~2(a) is the topologically same as 
that in Ref.{\cite{KLR}}.
This confirms qualitative feature of our calculation.
Major differences between our results and those of Ref.{\cite{KLR}} are 
the transition temperature of our mean-field calculation being 
as much as twice of that in Ref.{\cite{KLR}} and existence of 
the "magnetic" phase $\langle u00 \rangle$ phase at low temperatures.

As shown in Fig.~2(b), $\langle uu00 \rangle$ phase corresponding
to the ground state structure in which two out of four rows of metal atoms
are missing appears for a certain range of chemical potential at low
temperatures. P.~J.~Kundrotas {\it et al\/} {\cite{KLR}} did 
not take this $\langle uu00 \rangle$ into account.
From predictions, the modulated phase accompanied 
by the wave number $Q=\cos^{-1}{(-K_1/4K_2)}$ should 
appear in this case. However, we could not find it because
our calculation is limited within the periodicity 12 and
not enough big to find it.

Fig.2(c) shows the phase diagram for the case c) where the condition 
$K_2/K_1 > 1/4\,\,\, (K_2 > 0)$ is satisfied. 
Therefore, the modulated phase with the wave number 
$Q=\cos^{-1}{(-K_1/4K_2)}\simeq Q_1=10\pi/12$
should appear in this case.
Actually, we obtain  within repeating unit 12 the stable modulated phase 
shown in Fig.3 in the 
vicinity of the  point ($\mu\,=\,0$ , $T_y(Q_1)$). 
In the figure, $r$ represents the row numbers along the $[001]$
direction on the FCC(110) surface and $y_r=2-3c_r$ where $c_r$ describes
the probability that a metal atom row exists at the $r$th row. In comparison
with the structure $\langle uu0 \rangle$ at lower temperatures in the case
(shown in Fig.2(c)), the distribution of metal atom rows along the $[001]$
direction on the FCC(110) metal surface is modulated.
Thus our numerical results are consistent with the predictions. 
Although it will be 
difficult to find out experimentally the modulated phase, 
we show a possibility of  realization of the modulated phases.

%
The details of the present calculation  will be published near future

\begin{figure}[t]
\begin{center}
\noindent
\epsfxsize=0.45\textwidth
\epsfbox{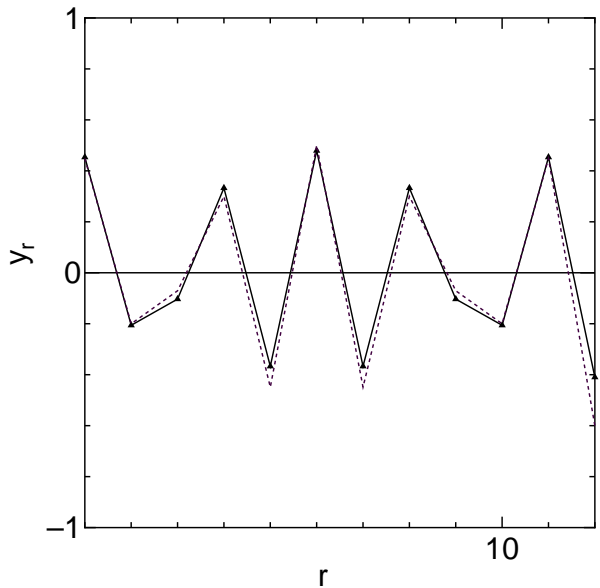}
\end{center}
\caption{$r$-dependence of $y_{r}$ in the modulated phase. The dotted line
represents $0.6\cos(Q_1(r-6))-0.1\cos(Q_2(r-6))$, where $Q_1=10\pi/12$, and
$Q_2=9\pi/12$.
}
\end{figure}

\section{Summary}
The ground state and finite temperature phase diagrams of 
adsorbate-induced row-type-alignments
is calculated by the Blume-Emmery-Griffiths (BEG) model. 
In the model, we introduce adatom-adatom and dipole-dipole interactions 
between nearest-neighbor (NN) and next-nearest-neighbor (NNN) rows. 

The ground state analysis shows the new structure corresponding to the 
$\langle uudd \rangle$ structure of the ANNNI model appears under 
the certain conditions for  NNN interactions. 

For the finite temperature,  we obtain 
the condition for the interactions to realize the stable 
modulated phases by solving the linearized 
mean-field equations analytically. 

 Numerical analysis of the mean-field equations 
is performed to calculate the finite temperature phase diagrams for 
several interaction regimes, including the region satisfying 
the modulated phase conditions. These analyses show that 
there appear coverage modulated  (incommensurate)  
phases at high temperatures for certain regime of the interactions.

\end{document}